\documentclass[12pt]{article}
\usepackage{amssymb}

\input{tcilatex}

\begin{document}

\smallskip\ 

\begin{center}
\textbf{SUPERFIELD DESCRIPTION OF A SELF-DUAL}

\smallskip\ 

\textbf{SUPERGRAVITY a la MACDOWELL-MANSOURI}

\textbf{\ }

\smallskip\ 

J. A. Nieto \footnote{%
nieto@uas.uasnet.mx}

\smallskip

\textit{Facultad de Ciencias F\'{\i}sico-Matem\'{a}ticas, Universidad Aut%
\'{o}noma}

\textit{de Sinaloa, C.P. 80000, Culiac\'{a}n Sinaloa, M\'{e}xico}

\bigskip\ 

\bigskip\ 

\textbf{Abstract}
\end{center}

Using MacDowell-Mansouri theory, in this work, we investigate a superfield
description of the self-dual supergravity \textit{a la} Ashtekar. We find
that in order to reproduce previous results on supersymmetric Ashtekar
formalism, it is necessary to properly combine the supersymmetric
field-strength in the Lagrangian. We extend our procedure to the case of
supersymmetric Ashtekar formalism in eight dimensions.

\bigskip\ 

\bigskip\ 

\bigskip\ 

\bigskip\ 

\bigskip\ 

\bigskip\ 

\bigskip\ 

Keywords: self-dual gravity, MacDowell-Mansouri, Ashtekar theory

Pacs numbers: 04.60.-m, 04.65.+e, 11.15.-q, 11.30.Ly

March, 2006

\newpage \noindent \textbf{1.- Introduction}

\smallskip\ 

The MacDowell-Mansouri formalism [1]-[2] (see also Refs. [3]-[6]) is one of
the most interesting attempts to describe gravity as a gauge field theory.
The basic assumption in such a formalism is to consider a curvature defined
in terms of a $SO(1,4)-$valued one-form connection $A^{AB}$ which is broken
into a $SO(1,3)-$valued one-form connection $A^{ab}$ and one-form tetrad
field $A^{4a}=e^{a}.$ The corresponding proposed action is quadratic in the
reduced action with the skew symmetric $\varepsilon -$symbol acting as a
contracting object. An interesting aspect of this action is that it is
reduced to the Einstein-Hilbert action when the $\varepsilon -$symbol is
associated with a four dimensional spacetime. Indeed, since a curvature is a
two-form the quadratic action gives a four-form, which can be contracted
with an $\varepsilon -$symbol of the form $\varepsilon ^{\mu \nu \alpha
\beta }$. These observations mean that the MacDowell-Mansouri theory is
intrinsically a four dimensional gauge theory of gravity. On the other hand,
one of the most interesting alternatives for quantum gravity is provided by
the Ashtekar theory [7] which is also intrinsically a four dimensional
theory of gravity. So a natural step further was to see if the
MacDowell-Mansouri formalism and the Ashtekar theory were related. It turns
out that a connection between these two theories was achieved by Nieto,
Socorro, and Obreg\'{o}n [8] whom proposed a MacDowell-Mansouri type action,
but with the curvature replaced by a self-dual curvature. The interesting
thing of this approach is that by dropping the Euler and Pontrjagin
topological invariants the Nieto-Socorro-Obreg\'{o}n action leads to the
Ashtekar formalism via the Jacobson-Smolin-Samuel action [9].

Moreover, it has been shown that the Nieto-Socorro-Obreg\'{o}n action can be
extended to the case of a self-dual supergravity in four dimensions [10].
However, this treatment refers only to the case of pure supergravity. If one
desires to introduce interactions some problems arise -similar to the ones
presented in the original theory of the MacDowell-Mansouri theory for
supergravity. In particular, the need of introducing a metric of the
spacetime weakens the original idea of seeing the self-dual supergravity as
a gauge theory. Thus, one is forced to look for an alternative for
supersymmetrizing the self-dual gravity with the hope that the interactions
appear more naturally and in the spirit of a gauge theory.

As it is known, a superfield formalism [11] provides one of best mechanisms
to supersymmetrisize a physical system, and is a particularly useful tool in
the construction of interacting Lagrangians. In this work we explore the
possibility of making a description of the self-dual supergravity a la
MacDowell-Mansouri theory in terms of the superfield formalism. We prove
that our description leads exactly to the same action as the one proposed by
Nieto-Socorro-Obreg\'{o}n for the case of the self-dual supergravity in four
dimensions.

Another source of motivation for the present work may arise from the
following observations. Traditionally, the fact that the Ashtekar theory
requires a four dimensional scenario for its formulation, it is seen as a
key feature over the alternative for quantum gravity based on a superstring
theory which requires a higher dimensional spacetime background. However, if
such Jacobson-Samuel-Smolin action is considered as a BF theory, then a
possible generalization to higher dimensions of the Ashtekar formalism could
be possible [12]. But in this BF approach the self-duality concept is lost.
An alternative has been proposed by Nieto [13] in order to promote the
Ashtekar formalism in the sense of self-duality of higher dimensions. In
this case, the octonion structure [14]-[16] becomes the mathematical key
tool. In fact, by using an octonion algebra in Ref. [13] it has been shown
that it makes sense to consider an Ashtekar formalism in eight dimensions.
It turns out that this generalization opens the possibility to consider
Ashtekar formalism in twelve dimensions either by the reduction $%
10+2\rightarrow (1+3)+(1+7)$ or by the prescription $10+2\rightarrow
(2+2)+(0+8)$ [17]. The case $(1+3)+(1+7)$ is particularly interesting
because establishes that the Ashtekar formalism is not completely unrelated
to the string theory or $M-$theory. While the case $(2+2)+(0+8)$ is
interesting because $(2+2)$ and $(0+8)$ are exceptional signatures. Both
proposals $10+2\rightarrow (1+3)+(1+7)$ and $10+2\rightarrow (2+2)+(0+8)$
have been developed using only a bosonic degree of freedom and it is the
goal of this work to direct the first steps toward a supersymmetrization of
such proposals.

Technically, the article is organized as follows: In section 2, we give a
short sketch of the self-dual MacDowell-Mansouri theory. In section 3, we
briefly describe the supersymmetric MacDowell-Mansouri theory. In section 4,
we explicitly show that such a supersymmetric version of the
MacDowell-Mansouri theory can be accomplished via superfield formalism.
Finally, in section 5, we outline a possible application of our results to
the case of an Ashtekar formalism in a spacetime of $2+10-$dimensional
spacetime.

\bigskip\ 

\noindent \textbf{2.- Self-dual MacDowell-Mansouri theory}

\smallskip\ 

The theory of MacDowell-Mansouri is a pure gauge theory in $3+1$ dimensions,
with the anti-de Sitter group $SO(3,2)$ as the gauge group. After breaking
the original gauge group to $SO(3,1)$, the resultant $SO(3,1)$ gauge theory
leads to the Einstein-Hilbert action, the cosmological constant term and the
Euler topological invariant.

By convenience, let us recall MacDowell-Mansouri mechanism. Consider the
MacDowell-Mansouri type action in a $1+3-$dimensional spacetime $M^{1+3},$

\begin{equation}
S=\int_{M^{1+3}}d^{4}x\varepsilon ^{\mu \nu \rho \sigma }\mathcal{R}_{\mu
\nu }^{~ab}\mathcal{R}_{\rho \sigma }^{~cd}\varepsilon _{abcd}.  \tag{1}
\end{equation}%
Here, $\varepsilon ^{\mu \nu \rho \sigma }$ and $\varepsilon _{abcd}$ are
the completely antisymmetric $\varepsilon $-symbols and $\mathcal{R}_{\mu
\nu }^{~ab}$ is given by%
\begin{equation}
\mathcal{R}_{\mu \nu }^{~ab}=F_{\mu \nu }^{~ab}+\Sigma _{\mu \nu }^{~ab}, 
\tag{2}
\end{equation}%
with\ 

\begin{equation}
F_{\mu \nu }^{~ab}=\partial _{\mu }A_{\nu }^{~ab}-\partial _{\nu }A_{\mu
}^{~ab}+A_{\mu }^{~ac}A_{\nu c}^{~b}-A_{\mu }^{~bc}A_{\nu c}^{~a}  \tag{3}
\end{equation}%
and

\begin{equation}
\Sigma _{\mu \nu }^{~ab}=e_{\mu }^{a}e_{\nu }^{b}-e_{\mu }^{b}e_{\nu }^{a}. 
\tag{4}
\end{equation}%
where $A_{\mu }^{~ab}$ is a $SO(1,3)$-connnection and $e_{\mu }^{a}$ is a
tetrad field.

Substituting (2) into (1) we get

\begin{equation}
S=\int_{M^{1+3}}(T+K+C),  \tag{5}
\end{equation}%
with

\begin{equation}
T=\varepsilon ^{\mu \nu \alpha \beta }F_{\mu \nu }^{~ab}F_{\alpha \beta
}^{~cd}\varepsilon _{abcd},  \tag{6}
\end{equation}

\begin{equation}
K=2\varepsilon ^{\mu \nu \alpha \beta }\Sigma _{\mu \nu }^{~ab}F_{\alpha
\beta }^{~cd}\varepsilon _{abcd},  \tag{7}
\end{equation}%
and

\begin{equation}
C=\varepsilon ^{\mu \nu \alpha \beta }\Sigma _{\mu \nu }^{~ab}\Sigma
_{\alpha \beta }^{~cd}\varepsilon _{abcd}.  \tag{8}
\end{equation}%
It is not difficult to recognize that $T$ leads to the Euler topological
invariant, $K$ determines the Einstein-Hilbert action and $C$ gives the
cosmological constant term.

In the case of the self-dual (antiself-dual) MacDowell-Mansouri theory the
action (1) is extended in the following form

\begin{equation}
S=\int_{M^{1+3}}d^{4}x\varepsilon ^{\mu \nu \rho \sigma +}\mathcal{R}_{\mu
\nu }^{~ab+}\mathcal{R}_{\rho \sigma }^{~cd}\varepsilon _{abcd},  \tag{9}
\end{equation}%
where

\begin{equation}
^{\pm }\mathcal{R}_{\mu \nu }^{~ab}=\frac{1}{2}^{\pm }B_{cd}^{ab}\mathcal{R}%
_{\mu \nu }^{~cd}.  \tag{10}
\end{equation}%
Here,

\begin{equation}
^{\pm }B_{cd}^{ab}=\frac{1}{2}(\delta _{cd}^{ab}\pm i\varepsilon _{cd}^{ab}),
\tag{11}
\end{equation}%
where $\delta _{cd}^{ab}=\delta _{c}^{a}\delta _{d}^{b}-\delta
_{d}^{a}\delta _{c}^{b}$ is a generalized delta.

We can again write (9) as in (5), namely

\begin{equation}
S^{\prime }=\int_{M^{1+3}}(T^{\prime }+K^{\prime }+C^{\prime }),  \tag{12}
\end{equation}%
where $T^{\prime },K^{\prime }$ and $C^{\prime }$ have the same form as $T,K$
and $C$, but with $F_{\mu \nu }^{~ab}$ and $\Sigma _{\mu \nu }^{~ab}$
replaced by $^{+}F_{\mu \nu }^{~ab}$ and $^{+}\Sigma _{\mu \nu }^{~ab},$
respectively.

Using the identity

\begin{equation}
^{+}B_{ef}^{ab+}B_{gh}^{cd}\varepsilon _{abcd}=2(\varepsilon _{efgh}-i\delta
_{efgh}),  \tag{13}
\end{equation}%
with $\delta _{efgh}=\eta _{eg}\eta _{fh}-\eta _{eh}\eta _{fg}$, we discover
that $T^{\prime },K^{\prime }$ and $C^{\prime }$ can be written as

\begin{equation}
T^{\prime }=\frac{1}{2}(\varepsilon ^{\mu \nu \alpha \beta }F_{\mu \nu
}^{~ab}F_{\alpha \beta }^{~cd}\varepsilon _{abcd}-i\varepsilon ^{\mu \nu
\alpha \beta }F_{\mu \nu }^{~ab}F_{\alpha \beta }^{~cd}\delta _{abcd}), 
\tag{14}
\end{equation}

\begin{equation}
K^{\prime }=(\varepsilon ^{\mu \nu \alpha \beta }\Sigma _{\mu \nu
}^{~ab}F_{\alpha \beta }^{~cd}\varepsilon _{abcd}-i\varepsilon ^{\mu \nu
\alpha \beta }\Sigma _{\mu \nu }^{~ab}F_{\alpha \beta }^{~cd}\delta _{abcd}),
\tag{15}
\end{equation}%
and

\begin{equation}
C^{\prime }=\frac{1}{2}(\varepsilon ^{\mu \nu \alpha \beta }\Sigma _{\mu \nu
}^{~ab}\Sigma _{\alpha \beta }^{~cd}\varepsilon _{abcd}-i\varepsilon ^{\mu
\nu \alpha \beta }\Sigma _{\mu \nu }^{~ab}\Sigma _{\alpha \beta
}^{~cd}\delta _{abcd}).  \tag{16}
\end{equation}%
The first and the second term in $T^{\prime }$ corresponds to the Euler and
Pontrjagin topological invariants respectively. The first term in $K^{\prime
}$ leads to (7), while the second term in (15) vanishes identically after
using the cyclical identities of $F_{\alpha \beta }^{~cd}$. Similarly, the
first term in $C^{\prime }$ corresponds to (8), while the second term of $%
C^{\prime }$ is identically zero. These results show that up to the
Pontrjagin topological invariant the actions (1) and (9) are equivalents. It
is worth observing that, despite $^{\pm }\mathcal{R}_{\mu \nu }^{~ab}$ is a
complex field, these results imply that, classically, the action (9) itself
can be reduced to a real quantity (see Refs. [8]-[10] for details). At the
quantum level, however, this is not true, but one can use Ashtekar mechanism
[7] in order to develop a consistent canonical quantization out of the
action (9).

\bigskip\ 

\noindent \textbf{3.- Supersymmetric MacDowell-Mansouri theory}

\smallskip\ 

The supersymmetric version of MacDowell-Mansouri theory can be simply
constructed by promoting the $SO(3,2)$ gauge fields to the corresponding
supergroup $OSp(1|4)$ gauge theory. In particular, the gauge potential $%
\mathcal{A}_{\mu }^{IJ}$ is now seen as a Lie algebra $osp(1|4)$-valued
potential. The corresponding field strength $\mathcal{F}_{\mu \nu }^{~IJ}$
is given by

\begin{equation}
\mathcal{F}_{\mu \nu }^{~IJ}=\partial _{\mu }\mathcal{A}_{\nu
}^{~IJ}-\partial _{\nu }\mathcal{A}_{\mu }^{~IJ}+\frac{1}{2}f_{JKLM}^{IJ}%
\mathcal{A}_{\mu }^{~JK}\mathcal{A}_{\nu }^{~LM},  \tag{17}
\end{equation}%
where $f_{JKLM}^{IJ}$ are the structure constants of the super Lie algebra $%
osp(1|4)$. The field strength $\mathcal{F}_{\mu \nu }^{~IJ}$ can be
decomposed in three terms corresponding to the three generators $%
S_{IJ}=(S_{ab},P_{a},Q_{i})$ (with $P_{a}=S_{4a}$) of $osp(1|4)$ as (see
[18]-[19] and references therein)

\begin{equation}
\mathcal{F}_{\mu \nu }^{ab}=F_{\mu \nu }^{ab}+\Sigma _{\mu \nu }^{ab}+\Psi
_{\mu \nu }^{ab},\ \ \ \mathcal{F}_{\mu \nu }^{i}=F_{\mu \nu }^{i}+\Sigma
_{\mu \nu }^{i},\ \ \ \mathcal{F}_{\mu \nu }^{a}=F_{\mu \nu }^{a}+\Sigma
_{\mu \nu }^{a},  \tag{18}
\end{equation}%
where

\begin{equation}
F_{\mu \nu }^{~ab}=\partial _{\mu }A_{\nu }^{~ab}-\partial _{\nu }A_{\mu
}^{~ab}+\frac{1}{2}f_{cdef}^{ab}A_{\mu }^{~cd}A_{\nu }^{~ef},  \tag{19}
\end{equation}%
\begin{equation}
\Sigma _{\mu \nu }^{ab}=2f_{4c4d}^{ab}A_{\mu }^{~4c}A_{\nu }^{~4d},  \tag{20}
\end{equation}%
\begin{equation}
\Psi _{\mu \nu }^{ab}={\frac{1}{2}}f_{ij}^{ab}A_{\mu }^{~i}A_{\nu }^{~j}, 
\tag{21}
\end{equation}%
\begin{equation}
\Sigma _{\mu \nu }^{a}={\frac{1}{2}}f_{ij}^{4a}A_{\mu }^{~i}A_{\nu }^{~j}, 
\tag{22}
\end{equation}%
\begin{equation}
\Sigma _{\mu \nu }^{i}=f_{4aj}^{i}(A_{\mu }^{~4a}A_{\nu }^{~j}-A_{\nu
}^{~4a}A_{\mu }^{~j}),  \tag{23}
\end{equation}%
and%
\begin{equation}
F_{\mu \nu }^{~i}=\partial _{\mu }A_{\nu }^{~i}-\partial _{\nu }A_{\mu }^{i}+%
\frac{1}{2}f_{cdj}^{i}(A_{\mu }^{~cd}A_{\nu }^{~j}-A_{\nu }^{~cd}A_{\mu
}^{~j}).  \tag{24}
\end{equation}%
Consider the action [10]

\begin{equation}
S=\int d^{4}x\varepsilon ^{\mu \nu \rho \sigma }\mathcal{F}_{\mu \nu }^{~~A}%
\mathcal{F}_{\rho \sigma }^{~~B}g_{AB},  \tag{25}
\end{equation}%
where $g_{AB}$ is the invariant diagonal metric in $osp(1|4)$ and it is
defined by 
\begin{equation}
g_{AB}=\left( 
\begin{array}{cc}
\varepsilon _{abcd} & 0 \\ 
0 & (C\Gamma _{5})_{ij}%
\end{array}%
\right) ,  \tag{26}
\end{equation}%
where $\Gamma _{5}=i\Gamma _{0}\Gamma _{1}\Gamma _{2}\Gamma _{3}$ and $%
\Gamma _{\mu }$ are the Dirac matrices satisfying the Clifford algebra $%
\{\Gamma _{\mu },\Gamma _{v}\}=-2\eta _{\mu v},$ with $\eta _{\mu
v}=diag(-1,1,1,1)$. In terms of its metric components the above-mentioned
action can be written as [10], [19]

\begin{equation}
S=\int d^{4}x\varepsilon ^{\mu \nu \rho \sigma }\bigg(\mathcal{F}_{\mu \nu
}^{~ab}\mathcal{F}_{\rho \sigma }^{~cd}\varepsilon _{abcd}+\mathcal{F}_{\mu
\nu }^{~i}\mathcal{F}_{\rho \sigma }^{~j}(C\Gamma _{5})_{ij}\bigg).  \tag{27}
\end{equation}%
By identifying $A_{\mu }^{~ab}\equiv \omega _{\mu }^{~ab}$ with the spin
connection, $A_{\mu }^{~4a}\equiv e_{\mu }^{~a}$ with the tetrad and $A_{\mu
}^{~i}\equiv \psi _{\mu }^{~i}$ with the gravitino field we discover that
the action (27) results in the gauge theory of $N=1$ supergravity (see Ref.
[10] for details).

\bigskip\ 

\noindent \textbf{4- Superfield formalism of self-dual MacDowell-Mansouri}

\smallskip\ 

In order to obtain the self-dual part of (27), we shall combine the
superfield formalism with the internal supergroup $OSp(1|4).$ Let us
consider the supersymmetric field-strengths [11]

\begin{equation}
W_{(\alpha )}=-i\lambda _{(\alpha )}+\theta _{(\alpha )}D-\sigma
_{\;\;\;(\alpha )}^{\mu \upsilon \;(\beta )}\theta _{(\beta )}\mathcal{F}%
_{\mu \nu }+\theta ^{2}\sigma _{(\alpha \dot{\beta})}^{\mu }\nabla _{\mu }%
\bar{\lambda}^{(\dot{\beta})},  \tag{28}
\end{equation}

\begin{equation}
W_{{}}^{(\alpha )}=-i\lambda _{{}}^{(\alpha )}+\theta ^{(\alpha )}D+\theta
^{(\beta )}\sigma _{\;\;\;(\beta )}^{\mu \nu \;(\alpha )}\mathcal{F}_{\mu
\nu }-\theta ^{2}\bar{\sigma}_{{}}^{\mu (\dot{\beta}\alpha )}\nabla _{\mu }%
\bar{\lambda}_{(\dot{\beta})}^{{}},  \tag{29}
\end{equation}

\begin{equation}
\bar{W}_{{}}^{(\dot{\alpha})}=i\bar{\lambda}_{{}}^{(\dot{\alpha})}+\bar{%
\theta}^{(\dot{\alpha})}D+\bar{\sigma}_{\qquad (\dot{\beta})}^{\mu \nu (\dot{%
\alpha})}\bar{\theta}^{(\dot{\beta})}\mathcal{F}_{\mu \nu }-\bar{\theta}^{2}%
\bar{\sigma}_{{}}^{\mu (\dot{\alpha}\alpha )}\nabla _{\mu }\lambda _{(\alpha
)},  \tag{30}
\end{equation}%
and

\begin{equation}
\bar{W}_{(\dot{\alpha})}^{{}}=i\bar{\lambda}_{(\dot{\alpha})}^{{}}+\bar{%
\theta}_{(\dot{\alpha})}^{{}}D-\bar{\theta}_{(\dot{\beta})}\bar{\sigma}%
_{\qquad (\dot{\alpha})}^{\mu \nu (\dot{\beta})}\mathcal{F}_{\mu \nu }+\bar{%
\theta}^{2}\sigma _{(\alpha \dot{\alpha})}^{\mu }\nabla _{\mu }\lambda
_{{}}^{(\alpha )},  \tag{31}
\end{equation}%
where $(\alpha ),(\beta )$ $etc.$ are spinor indices, while $\mu ,\nu $ $etc$
are space-time indices. (At this moment, we closely follow the References in
[11].) Using these expressions we find

\begin{equation}
W^{(\alpha )}W_{(\alpha )}\mid _{\theta \theta }=-\frac{1}{2}\mathcal{F}%
^{\mu \nu }\mathcal{F}_{\mu \nu }-\frac{i}{4}\varepsilon ^{\mu \nu \rho
\sigma }\mathcal{F}_{\mu \nu }\mathcal{F}_{\rho \sigma }-2i\lambda ^{(\alpha
)}\sigma _{(\alpha \dot{\beta})}^{\mu }\nabla _{\mu }\bar{\lambda}^{(\dot{%
\beta})}+D^{2},  \tag{32}
\end{equation}%
and as well as

\begin{equation}
\bar{W}_{(\dot{\alpha})}\bar{W}_{{}}^{(\dot{\alpha})}\mid _{\bar{\theta}\bar{%
\theta}}=-\frac{1}{2}\mathcal{F}^{\mu \nu }\mathcal{F}_{\mu \nu }+\frac{i}{4}%
\varepsilon ^{\mu \nu \rho \sigma }\mathcal{F}_{\mu \nu }\mathcal{F}_{\rho
\sigma }-2i\lambda ^{(\alpha )}\sigma _{(\alpha \dot{\beta})}^{\mu }\nabla
_{\mu }\bar{\lambda}^{(\dot{\beta})}+D^{2}+\nabla _{\mu }J^{\mu },  \tag{33}
\end{equation}%
where $J^{\mu }=2i\lambda \sigma ^{\mu }\bar{\lambda}.$

Therefore, we get

\begin{equation}
L=\frac{1}{4}(W^{(\alpha )}W_{(\alpha )}\mid _{\theta \theta }+\bar{W}_{(%
\dot{\alpha})}\bar{W}_{{}}^{(\dot{\alpha})}\mid _{\bar{\theta}\bar{\theta}%
})=-\frac{1}{4}\mathcal{F}^{\mu \nu }\mathcal{F}_{\mu \nu }-i\lambda
^{(\alpha )}\sigma _{(\alpha \dot{\beta})}^{\mu }\nabla _{\mu }\bar{\lambda}%
^{(\dot{\beta})}+\frac{1}{2}D^{2},  \tag{34}
\end{equation}%
and also

\begin{equation}
L_{t}=\frac{i}{4}(W^{(\alpha )}W_{(\alpha )}\mid _{\theta \theta }-\bar{W}_{(%
\dot{\alpha})}\bar{W}_{{}}^{(\dot{\alpha})}\mid _{\bar{\theta}\bar{\theta}})=%
\frac{1}{8}\varepsilon ^{\mu \nu \rho \sigma }\mathcal{F}_{\mu \nu }\mathcal{%
F}_{\rho \sigma }.  \tag{35}
\end{equation}%
Here, we dropped the surface term $\nabla _{\mu }J^{\mu }$.

If we introduce the definition

\begin{equation}
{^{\pm }}\mathcal{F}_{\mu \nu }^{~~A}=\frac{1}{2}{^{\pm }}\mathcal{B}_{B}^{A}%
\mathcal{F}_{\mu \nu }^{B},  \tag{36}
\end{equation}%
where ${^{\pm }}\mathcal{B}_{B}^{A}$ is given by

\begin{equation}
\begin{array}{l}
{^{\pm }}\mathcal{B}_{B}^{A}=\left( 
\begin{array}{cc}
^{\pm }B_{cd}^{ab} & 0 \\ 
0 & ^{\pm }B_{j}^{i}%
\end{array}%
\right) \\ 
\\ 
=\left( 
\begin{array}{cc}
\delta _{cd}^{ab}\pm i\varepsilon _{cd}^{ab} & 0 \\ 
0 & \frac{1}{2}(1\pm \Gamma _{5})_{j}^{i}%
\end{array}%
\right) .%
\end{array}
\tag{37}
\end{equation}%
It is not difficult to see that the self-dual sector of $L_{t}$ is given by

\begin{equation}
^{+}L_{t}=\frac{i}{4}(^{+}W^{(\alpha )+}W_{(\alpha )}\mid _{\theta \theta
}-^{+}\bar{W}_{(\dot{\alpha})}^{+}\bar{W}_{{}}^{(\dot{\alpha})}\mid _{\bar{%
\theta}\bar{\theta}})=\frac{1}{8}\varepsilon ^{\mu \nu \rho \sigma +}%
\mathcal{F}_{\mu \nu }^{A~+}\mathcal{F}_{\mu \nu }^{B}g_{AB},  \tag{38}
\end{equation}%
where $g_{AB}$ is the metric associated with the superalgebra $osp(1|4)$ and
it is given by the formula (26).

Using (26) we discover that (38) can be written as:

\begin{equation}
^{+}L_{t}=\frac{1}{8}\varepsilon ^{\mu \nu \rho \sigma +}(^{+}\mathcal{F}%
_{\mu \nu }^{ab~+}\mathcal{F}_{\rho \sigma }^{cd}\varepsilon _{abcd}+^{+}%
\mathcal{F}_{\mu \nu }^{~i~+}\mathcal{F}_{\rho \sigma }^{j}(C\Gamma
_{5})_{ij}).  \tag{39}
\end{equation}%
In this expression, we recognize the Lagrangian proposed by Nieto, Socorro
and Obreg\'{o}n [10], which generalizes the corresponding Lagrangian (27) to
the self-dual case.

A generalization of (39) can be achieved by means of the action [19]

\begin{equation}
S=\int d^{4}x\varepsilon ^{\mu \nu \rho \sigma }\bigg({^{+}}\tau {^{+}}%
\mathcal{F}_{\mu \nu }^{A}{^{+}}\mathcal{F}_{\rho \sigma }^{B}g_{AB}-{^{-}}%
\tau {^{-}}\mathcal{F}_{\mu \nu }^{A}{^{-}}\mathcal{F}_{\rho \sigma
}^{B}g_{AB}\bigg),  \tag{40}
\end{equation}%
which considers both the self-dual sector as well as the antiself-dual
sector. In this part, ${^{+}}\tau $ and ${^{-}}\tau $ are constant coupling
parameters. The action (40) may provide the starting point for studying the
supersymmetric case of a $S-$duality program for gravity, which was
initiated in [19] for the MacDowell-Mansouri pure gauge theory.

\bigskip\ 

\noindent \textbf{5. Toward an eight dimensional superfield formalism of a
self-dual supergravity a la MacDowell-Mansouri }

\smallskip\ 

Now, we shall consider that it is possible to extend the procedure of
section 3 to eight-dimensions. Observe first the important role played by
the $\varepsilon -$symbols $\varepsilon ^{\mu \nu \rho \sigma }$, $%
\varepsilon ^{abcd}$ and $\Gamma _{5}$ in the expressions (37) and (38). In
fact, the symbol $\varepsilon ^{\mu \nu \rho \sigma }$ determines that the
dimensionality of the spacetime should be four; $\varepsilon ^{abcd}$ is $%
SO(3,1)-$invariant object and is used to define self-duality
(antiself-duality) in the bosonic sector, while $\Gamma _{5}=i\Gamma
_{0}\Gamma _{1}\Gamma _{2}\Gamma _{3}$ is directly related to the existence
of Dirac matrices satisfying a Clifford algebra in four dimensions and
determines the self-dual (antiself-dual) in the fermionic sector. In
general, since in any higher dimensional spacetime the field strength $%
\mathcal{F}_{\mu \nu }^{A}$ is a two-form matrix , it seems a difficult task
to consider its self-dual and antiself-dual sectors other than four
dimensions. Nevertheless, in the case of spin $1$ it has been proved that an
interesting possibility arises in eight dimensions [20]. The idea can be
traced back to the observation that $\varepsilon ^{0bcd}$ is linked to an
exceptional algebra: the quaternionic algebra. So, if one considers the
octonion structure, which is also an exceptional division algebra, one
should be able to consider the $\eta -$symbols $\eta ^{\mu \nu \rho \sigma }$%
, $\eta ^{abcd}$ and $\Gamma _{9}=i\Gamma _{0}\cdot \cdot \cdot \Gamma _{7}$%
. In fact, some progress in using the $\eta -$symbols has been achieved in
the self-dual Yang-Mills field case [20] (see also [21]) and in the bosonic
Ashtekar formalism in eight dimensions [13]. The key formula for applying
the $\eta -$symbols in these cases is

\begin{equation}
\eta _{\mu \nu \alpha \beta }\eta ^{\tau \sigma \alpha \beta }=6\delta _{\mu
\nu }^{\tau \sigma }+4\eta _{\mu \nu }^{\tau \sigma },  \tag{41}
\end{equation}%
which, in contrast to the analogue $\varepsilon -$symbol expression

\begin{equation}
\varepsilon _{\mu \nu \alpha \beta }\varepsilon ^{\tau \sigma \alpha \beta
}=2\delta _{\mu \nu }^{\tau \sigma },  \tag{42}
\end{equation}%
leads to self-duality of the field strength ${^{+}}\mathcal{F}_{\mu \nu
}^{ab}=\frac{1}{2}(\delta _{cd}^{ab}+\eta _{cd}^{ab})\mathcal{F}_{\mu \nu
}^{ab}$ in the form,

\begin{equation}
^{\ast +}\mathcal{F}_{\mu \nu }^{ab}=3^{+}\mathcal{F}_{\mu \nu }^{ab}. 
\tag{43}
\end{equation}%
Thus, except for a numerical factor $^{+}\mathcal{F}_{\mu \nu }^{ab}$
satisfies the usual self-dual relation.

In general, in a superspace in eight dimensions $(x^{\mu },\theta _{(A)}),$
with a $\mu =1,...,8$ and $(A)=1,...,16,$ a vector superfield can be written
as [22]

\begin{equation}
V_{J}=\tsum_{(A_{i})=1}^{16}\theta ^{(A_{1})}\cdot \cdot \cdot \theta
^{(A_{16})}V_{J(A_{1})\cdot \cdot \cdot (A_{16})}(x),  \tag{44}
\end{equation}%
where $\theta _{(A)}$ are Grassmann anticommuting variables. In order to
write the corresponding superfield strengths in eight dimensions, $%
W^{(\alpha )}$ and $\bar{W}_{(\dot{\alpha})}$, it turns out to be convenient
to work on the Euclidean sector and subsequent ones to perform a Wick
rotation. This allows to consider the $SO(8)$ matrices $\Gamma _{\mu },$
satisfying the Clifford algebra $\{\Gamma _{\mu },\Gamma _{v}\}=2\delta
_{\mu v},$ in the \textit{Spin}$(7)$ representation

\begin{equation}
\begin{array}{l}
\Gamma _{\hat{a}}=\left( 
\begin{array}{cc}
0 & -i\gamma _{\hat{a}} \\ 
i\gamma _{\hat{a}} & 0%
\end{array}%
\right) , \\ 
\\ 
\Gamma _{8}=\left( 
\begin{array}{cc}
0 & I \\ 
I & 0%
\end{array}%
\right) ,\;\;\Gamma _{9}\left( 
\begin{array}{cc}
I & 0 \\ 
0 & -I%
\end{array}%
\right) ,%
\end{array}
\tag{45}
\end{equation}%
where $(\gamma _{\hat{a}})_{(\alpha \beta )}$ $(\alpha ,\beta =1,...,8$ and $%
\hat{a}=1,...,7)$ are antisymmetric \textit{Spin}$(7)$ matrices defined by

\begin{equation}
\begin{array}{c}
(\gamma _{\hat{a}})_{(\hat{b}8)}=i\delta _{\hat{a}\hat{b}}, \\ 
\\ 
(\gamma _{\hat{a}})_{(\hat{b}\hat{c})}=iC_{\hat{a}\hat{b}\hat{c}}.%
\end{array}
\tag{46}
\end{equation}%
Here, $C_{\hat{a}\hat{b}\hat{c}}$ are the totally antisymmetric octonion
structure constants, which are related to the dual totally antisymmetric
octonionic tensor $\eta _{abcd}$ in form

\begin{equation}
C_{\hat{a}\hat{b}\hat{c}}=\eta _{8\hat{a}\hat{b}\hat{c}}  \tag{47}
\end{equation}%
and

\begin{equation}
^{\pm }F_{\hat{a}\hat{b}\hat{c}\hat{d}}\equiv C_{\hat{a}\hat{b}\hat{e}}C_{%
\hat{c}\hat{d}}^{\;\;\;\hat{e}}=\delta _{\hat{a}\hat{c}}\delta _{\hat{b}\hat{%
d}}-\delta _{\hat{a}\hat{d}}\delta _{\hat{b}\hat{c}}\pm \eta _{\hat{a}\hat{b}%
\hat{c}\hat{d}}.  \tag{48}
\end{equation}%
Observe that, since $\eta _{\hat{a}\hat{b}\hat{c}\hat{d}}$ is totally
antisymmetric, both solutions in (48) lead to the expression

\begin{equation}
C_{\hat{a}\hat{b}\hat{e}}C_{\hat{c}\hat{d}}^{\;\;\;\hat{e}}+C_{\hat{c}\hat{b}%
\hat{e}}C_{\hat{a}\hat{d}}^{\;\;\;\hat{e}}=2\delta _{\hat{a}\hat{c}}\delta _{%
\hat{b}\hat{d}}-\delta _{\hat{a}\hat{d}}\delta _{\hat{b}\hat{c}}-\delta _{%
\hat{c}\hat{d}}\delta _{\hat{b}\hat{a}},  \tag{49}
\end{equation}%
which determines the octonions as a normed algebra (see [23] and Refs.
therein).

For later computation we note that if one assumes the $^{+}F_{\hat{a}\hat{b}%
\hat{c}\hat{d}}$ solution then we obtain the identity

\begin{equation}
\eta _{\mu \nu \alpha \beta }\eta ^{\tau \sigma \alpha \beta }=6\delta _{\mu
\nu }^{\tau \sigma }+4\eta _{\mu \nu }^{\tau \sigma },  \tag{50}
\end{equation}%
while if one assumes the $^{-}F_{\hat{a}\hat{b}\hat{c}\hat{d}}$ solution we
get

\begin{equation}
\eta _{\mu \nu \alpha \beta }\eta ^{\tau \sigma \alpha \beta }=6\delta _{\mu
\nu }^{\tau \sigma }-4\eta _{\mu \nu }^{\tau \sigma }.  \tag{51}
\end{equation}

In the representation (45)-(46), the generators $\Gamma _{ab}=\frac{1}{2}%
[\Gamma _{a},\Gamma _{b}]$ of $SO(8)$ become

\begin{equation}
\begin{array}{l}
\Gamma _{\hat{a}\hat{b}}=\left( 
\begin{array}{cc}
(\gamma _{\hat{a}\hat{b}})_{(\alpha \beta )} & 0 \\ 
0 & (\gamma _{\hat{a}\hat{b}})_{(\alpha \beta )}%
\end{array}%
\right) , \\ 
\\ 
\Gamma _{\hat{a}8}=\left( 
\begin{array}{cc}
-i(\gamma _{\hat{a}})_{(\alpha \beta )} & 0 \\ 
0 & i(\gamma _{\hat{a}})_{(\alpha \beta )}%
\end{array}%
\right) .%
\end{array}
\tag{52}
\end{equation}%
Here, $(\gamma _{\hat{a}\hat{b}})_{(\alpha \beta )}$ are the antisymmetric 
\textit{Spin}$(7)$ generators,

\begin{equation}
\begin{array}{l}
(\gamma _{\hat{a}\hat{b}})_{(\hat{c}8)}=C_{\hat{a}\hat{b}\hat{c}}, \\ 
\\ 
(\gamma _{\hat{a}\hat{b}})_{(\hat{c}\hat{d})}=\delta _{\hat{a}\hat{c}}\delta
_{\hat{b}\hat{d}}-\delta _{\hat{a}\hat{d}}\delta _{\hat{b}\hat{c}}-\eta _{%
\hat{a}\hat{b}\hat{c}\hat{d}}.%
\end{array}
\tag{53}
\end{equation}

In order to split $\theta ^{(A)}$ into two Weyl spinors $\theta ^{(\alpha )}$
and $\bar{\theta}^{(\dot{\alpha})}$ one can use $\Gamma _{9}$ matrix$.$
Accordingly, using (45) we can write%
\begin{equation}
\begin{array}{l}
\Gamma _{a}=\left( 
\begin{array}{cc}
0 & \sigma _{a} \\ 
\bar{\sigma}_{a} & 0%
\end{array}%
\right) ,%
\end{array}
\tag{54}
\end{equation}%
where $\sigma _{a}=(-i\gamma _{\hat{a}},1)$ and $\bar{\sigma}_{a}=(i\gamma _{%
\hat{a}},1)$. Thus, we may introduce the corresponding \textit{Spin}$(7)$
generators

\begin{equation}
\sigma _{\;\;\;\;\;(\alpha )}^{ab\;(\beta )}=\frac{1}{2}(\sigma ^{a}\bar{%
\sigma}^{b}-\sigma ^{b}\bar{\sigma}^{a})_{\;\;(\alpha )}^{(\beta )}, 
\tag{55}
\end{equation}%
and

\begin{equation}
\bar{\sigma}_{\;\;\;(\alpha )}^{ab\;(\beta )}=\frac{1}{2}(\bar{\sigma}%
^{a}\sigma ^{b}-\bar{\sigma}^{b}\sigma ^{a})_{(\alpha )}^{\;(\beta )}. 
\tag{56}
\end{equation}%
In fact, using the $^{+}F_{\hat{a}\hat{b}\hat{c}\hat{d}}$ solution given in
(48) one may verify that

\begin{equation}
\begin{array}{l}
\sigma _{\hat{a}\hat{b}(\hat{c}8)}=C_{\hat{a}\hat{b}\hat{c}}, \\ 
\\ 
\sigma _{\hat{a}\hat{b}(\hat{c}\hat{d})}=\delta _{\hat{a}\hat{c}}\delta _{%
\hat{b}\hat{d}}-\delta _{\hat{a}\hat{d}}\delta _{\hat{b}\hat{c}}-\eta _{\hat{%
a}\hat{b}\hat{c}\hat{d}},%
\end{array}
\tag{57}
\end{equation}%
and also

\begin{equation}
\begin{array}{l}
\sigma _{\hat{a}8\hat{c}8}=\delta _{\hat{a}\hat{c}}, \\ 
\\ 
\sigma _{\hat{a}8\hat{b}\hat{c}}=C_{\hat{a}\hat{b}\hat{c}}.%
\end{array}
\tag{58}
\end{equation}%
We also get%
\begin{equation}
\begin{array}{l}
\bar{\sigma}_{\hat{a}\hat{b}(\hat{c}8)}=C_{\hat{a}\hat{b}\hat{c}}, \\ 
\\ 
\bar{\sigma}_{\hat{a}\hat{b}(\hat{c}\hat{d})}=\delta _{\hat{a}\hat{c}}\delta
_{\hat{b}\hat{d}}-\delta _{\hat{a}\hat{d}}\delta _{\hat{b}\hat{c}}-\eta _{%
\hat{a}\hat{b}\hat{c}\hat{d}},%
\end{array}
\tag{59}
\end{equation}%
and

\begin{equation}
\begin{array}{l}
\bar{\sigma}_{\hat{a}8\hat{c}8}=-\delta _{\hat{a}\hat{c}}, \\ 
\\ 
\bar{\sigma}_{\hat{a}8\hat{b}\hat{c}}=-C_{\hat{a}\hat{b}\hat{c}}.%
\end{array}
\tag{60}
\end{equation}

It is not difficult to see that the expressions (57) and (58) can be unified
in the form%
\begin{equation}
\sigma _{\mu \nu (\alpha \beta )}=\delta _{\mu \alpha }\delta _{\nu \beta
}-\delta _{\mu \beta }\delta _{\nu \alpha }-\eta _{\mu \nu \alpha \beta }. 
\tag{61}
\end{equation}%
If we use the identity (50) we find

\begin{equation}
\sigma _{\mu \nu (\lambda \tau )}\sigma _{\alpha \beta }^{\;\;(\lambda \tau
)}=8(\delta _{\mu \alpha }\delta _{\nu \beta }-\delta _{\mu \beta }\delta
_{\nu \alpha }).  \tag{62}
\end{equation}%
From the discussion of section 3, one should expect that in the computation
of $W^{(\alpha )}W_{(\alpha )}\mid _{\theta \theta }$ the expression (62)
leads to the analogue in eight dimensions of the term $-\frac{1}{2}\mathcal{F%
}^{\mu \nu }\mathcal{F}_{\mu \nu }$ rather than the analogue of the term $-%
\frac{1}{2}\mathcal{F}^{\mu \nu }\mathcal{F}_{\mu \nu }-\frac{i}{4}%
\varepsilon ^{\mu \nu \rho \sigma }\mathcal{F}_{\mu \nu }\mathcal{F}_{\rho
\sigma }.$ Moreover, we find that it seems to be no way to unified (59) and
(60) in just one expression for $\bar{\sigma}_{\mu \nu (\alpha \beta )}$ as
in the case of $\sigma _{\mu \nu (\alpha \beta )}$. These two observations
lead us to consider an alternative prescription.

Consider the quantities

\begin{equation}
\Theta _{\mu \nu \alpha \beta }=\delta _{\mu \alpha }\delta _{\nu \beta
}-\delta _{\mu \beta }\delta _{\nu \alpha }+\eta _{\mu \nu \alpha \beta } 
\tag{63}
\end{equation}%
and

\begin{equation}
\bar{\Theta}_{\mu \nu \alpha \beta }=\delta _{\mu \alpha }\delta _{\nu \beta
}-\delta _{\mu \beta }\delta _{\nu \alpha }-\eta _{\mu \nu \alpha \beta }, 
\tag{64}
\end{equation}%
which are related to the $^{+}F_{\hat{a}\hat{b}\hat{c}\hat{d}}$ and $^{-}F_{%
\hat{a}\hat{b}\hat{c}\hat{d}}$ solutions respectively. From the $^{+}F_{\hat{%
a}\hat{b}\hat{c}\hat{d}}$ solution we find that $\Theta _{\mu \nu \alpha
\beta }$ is a self-dual projector. In fact, using (50) we obtain

\begin{equation}
\Theta _{\mu \nu \tau \lambda }\Theta _{\alpha \beta }^{\tau \lambda
}=8\Theta _{\mu \nu \alpha \beta }  \tag{65}
\end{equation}%
and%
\begin{equation}
\frac{1}{2}\eta _{\mu \nu \tau \lambda }\Theta _{\alpha \beta }^{\tau
\lambda }=3\Theta _{\mu \nu \alpha \beta }.  \tag{66}
\end{equation}%
As it has been emphasized in Ref. [24] the object $\Theta _{\mu \nu \tau
\lambda }$ projects any antisymmetric second rank tensor onto its self-dual
parts \textbf{7}, according to the decomposition \textbf{28}=\textbf{7}$%
\oplus $\textbf{21} of the adjoint representation of $SO(8)\sim
SO(8)/Spin(7)\oplus Spin(7)$.

Similarly, using the formula (51) which corresponds to the $^{-}F_{\hat{a}%
\hat{b}\hat{c}\hat{d}}$ solution, we find that $\bar{\Theta}_{\mu \nu \tau
\lambda }$ satisfies

\begin{equation}
\bar{\Theta}_{\mu \nu \tau \lambda }\bar{\Theta}_{\alpha \beta }^{\tau
\lambda }=8\bar{\Theta}_{\mu \nu \alpha \beta }  \tag{67}
\end{equation}%
and

\begin{equation}
\frac{1}{2}\eta _{\mu \nu \tau \lambda }\bar{\Theta}_{\alpha \beta }^{\tau
\lambda }=-3\bar{\Theta}_{\mu \nu \alpha \beta }.  \tag{68}
\end{equation}%
Therefore $\bar{\Theta}_{\mu \nu \tau \lambda }$ is an antiself-dual
projector operator.

Thus, one should expect that in the computation of $W^{(\alpha )}W_{(\alpha
)}\mid _{\theta \theta }$ the expression (65) may lead to the analogue in
eight dimensions of the combination $-\frac{1}{2}\mathcal{F}^{\mu \nu }%
\mathcal{F}_{\mu \nu }-\frac{i}{4}\varepsilon ^{\mu \nu \rho \sigma }%
\mathcal{F}_{\mu \nu }\mathcal{F}_{\rho \sigma }$ in four dimensions. This
means that, for our purpose, the self-dual and the antiself-dual splitting
associated with the group $SO(8)$ provide a better alternative than the Weyl
splitting specified in (54).

With these tools at hand, one may proceed as in four dimensions, that is,
one may derive presumably from (44) the eight dimensional fields strengths

\begin{equation}
W_{(\alpha )}=-i\lambda _{(\alpha )}+\theta _{(\alpha )}D-\Theta
_{\;\;\;(\alpha )}^{\mu \upsilon \;(\beta )}\theta _{(\beta )}\mathcal{F}%
_{\mu \nu }+\theta ^{2}\sigma _{(\alpha \dot{\beta})}^{\mu }\nabla _{\mu }%
\bar{\lambda}^{(\dot{\beta})}+...,  \tag{69}
\end{equation}

\begin{equation}
W_{{}}^{(\alpha )}=-i\lambda _{{}}^{(\alpha )}+\theta ^{(\alpha )}D+\theta
^{(\beta )}\Theta _{\;\;\;(\beta )}^{\mu \nu \;(\alpha )}\mathcal{F}_{\mu
\nu }-\theta ^{2}\bar{\sigma}_{{}}^{\mu (\dot{\beta}\alpha )}\nabla _{\mu }%
\bar{\lambda}_{(\dot{\beta})}^{{}}+...,  \tag{70}
\end{equation}

\begin{equation}
\bar{W}_{{}}^{(\dot{\alpha})}=i\bar{\lambda}_{{}}^{(\dot{\alpha})}+\bar{%
\theta}^{(\dot{\alpha})}D+\bar{\Theta}_{\qquad (\dot{\beta})}^{\mu \nu (\dot{%
\alpha})}\bar{\theta}^{(\dot{\beta})}\mathcal{F}_{\mu \nu }-\bar{\theta}^{2}%
\bar{\sigma}_{{}}^{\mu (\dot{\alpha}\alpha )}\nabla _{\mu }\lambda _{(\alpha
)}...,  \tag{71}
\end{equation}%
and

\begin{equation}
\bar{W}_{(\dot{\alpha})}^{{}}=i\bar{\lambda}_{(\dot{\alpha})}^{{}}+\bar{%
\theta}_{(\dot{\alpha})}^{{}}D-\bar{\theta}_{(\dot{\beta})}\bar{\Theta}%
_{\qquad (\dot{\alpha})}^{\mu \nu (\dot{\beta})}\mathcal{F}_{\mu \nu }+\bar{%
\theta}^{2}\sigma _{(\alpha \dot{\alpha})}^{\mu }\nabla _{\mu }\lambda
_{{}}^{(\alpha )}+...  \tag{72}
\end{equation}%
Here, by convenience, we only wrote the relevant terms in the $\theta $
power expansion. The object $\mathcal{F}_{\mu \nu }$ should be seen now as
an eight dimensional nonabelian field strength and $\lambda _{(\alpha )}$ as
an eight dimensional chiral spinor.

Using (69)-(72) and (65) we find

\begin{equation}
W^{(\alpha )}W_{(\alpha )}\mid _{\theta \theta }=16\mathcal{F}^{\mu \nu }%
\mathcal{F}_{\mu \nu }+8\eta ^{\mu \nu \rho \sigma }\mathcal{F}_{\mu \nu }%
\mathcal{F}_{\rho \sigma }-2i\lambda ^{(\alpha )}\sigma _{(\alpha \dot{\beta}%
)}^{\mu }\nabla _{\mu }\bar{\lambda}^{(\dot{\beta})}+D^{2}.  \tag{73}
\end{equation}%
While if we use (67) we get

\begin{equation}
\bar{W}_{(\dot{\alpha})}\bar{W}_{{}}^{(\dot{\alpha})}\mid _{\bar{\theta}\bar{%
\theta}}=16\mathcal{F}^{\mu \nu }\mathcal{F}_{\mu \nu }-8\eta ^{\mu \nu \rho
\sigma }\mathcal{F}_{\mu \nu }\mathcal{F}_{\rho \sigma }-2i\lambda ^{(\alpha
)}\sigma _{(\alpha \dot{\beta})}^{\mu }\nabla _{\mu }\bar{\lambda}^{(\dot{%
\beta})}+D^{2},  \tag{74}
\end{equation}%
where we dropped the surface term $\nabla _{\mu }(2i\lambda \sigma ^{\mu }%
\bar{\lambda}).$ (It is worth mentioning that (73) corresponds to the
Lagrangian discussed in Ref. [24] for the case of $7D$ super Yang-Mills
theory.) Therefore, we get

\begin{equation}
L=\frac{1}{4}(W^{(\alpha )}W_{(\alpha )}\mid _{\theta \theta }+\bar{W}_{(%
\dot{\alpha})}\bar{W}_{{}}^{(\dot{\alpha})}\mid _{\bar{\theta}\bar{\theta}%
})=8\mathcal{F}^{\mu \nu }\mathcal{F}_{\mu \nu }-i\lambda ^{(\alpha )}\sigma
_{(\alpha \dot{\beta})}^{\mu }\nabla _{\mu }\bar{\lambda}^{(\dot{\beta})}+%
\frac{1}{2}D^{2},  \tag{75}
\end{equation}%
and also

\begin{equation}
L_{t}=\frac{1}{4}(W^{(\alpha )}W_{(\alpha )}\mid _{\theta \theta }-\bar{W}_{(%
\dot{\alpha})}\bar{W}_{{}}^{(\dot{\alpha})}\mid _{\bar{\theta}\bar{\theta}%
})=4\eta ^{\mu \nu \rho \sigma }\mathcal{F}_{\mu \nu }\mathcal{F}_{\rho
\sigma }.  \tag{76}
\end{equation}

If we now introduce the definition

\begin{equation}
{^{\pm }}\mathcal{F}_{\mu \nu }^{~~A}=\frac{1}{2}{^{\pm }}\mathcal{B}_{B}^{A}%
\mathcal{F}_{\mu \nu }^{B},  \tag{77}
\end{equation}%
where ${^{\pm }}\mathcal{B}_{B}^{A}$ is given by

\begin{equation}
\begin{array}{l}
{^{\pm }}\mathcal{B}_{B}^{A}=\left( 
\begin{array}{cc}
^{\pm }B_{cd}^{ab} & 0 \\ 
0 & ^{\pm }B_{j}^{i}%
\end{array}%
\right) \\ 
\\ 
=\left( 
\begin{array}{cc}
\delta _{cd}^{ab}\pm \eta _{cd}^{ab} & 0 \\ 
0 & \frac{1}{2}(1\pm \Gamma _{9})_{j}^{i}%
\end{array}%
\right) ,%
\end{array}
\tag{78}
\end{equation}%
we find the Lagrangian

\begin{equation}
^{+}L_{t}=\frac{1}{4}(^{+}W^{(\alpha )+}W_{(\alpha )}\mid _{\theta \theta
}-^{+}\bar{W}_{(\dot{\alpha})}^{+}\bar{W}_{{}}^{(\dot{\alpha})}\mid _{\bar{%
\theta}\bar{\theta}})=4\eta ^{\mu \nu \rho \sigma ~+}\mathcal{F}_{\mu \nu
}^{A~+}\mathcal{F}_{\mu \nu }^{B}g_{AB},  \tag{79}
\end{equation}%
which describes the self-dual sector of $L_{t}$ in eight dimensions. Of
course, in this case the metric $g_{AB}$ should be chosen as

\begin{equation}
g_{AB}=\left( 
\begin{array}{cc}
\eta _{abcd} & 0 \\ 
0 & (C\Gamma _{9})_{ij}%
\end{array}%
\right) .  \tag{80}
\end{equation}%
Therefore, we have derived a self-dual supergravity Lagrangian in eight
dimensions via superfield formalism. It turns out that, the bosonic sector
of the Lagrangian (79) is reduced to the Lagrangian proposed in the Ref.
[13]. This seems to be an important result in the quest of a supersymmetric
Ashtekar formalism in eight dimensions. Moreover, there is an important
observation that we need to make in connection with the invariance of
Lagrangian (79). In general the $\varepsilon $-symbol is Lorentz invariant
in any dimension, but in contrast the $\eta $-symbol is only $SO(7)$%
-invariant in eight dimensions (see Ref. [25]). Therefore, the $\eta $%
-symbol spoils the Lorentz invariance of the Lagrangian $^{+}L_{t}$ given in
(79), but maintains a hidden $SO(7)$-invariance. In fact, this is a general
phenomenon in field theories involving the $\eta $-symbol (see, for
instance, Refs. [24] and [27]).

\bigskip\ 

\noindent \textbf{6. Final remarks}

\smallskip\ 

It is well known the great importance of the role that a nonabelian field
strength plays in the gauge field theories. In a sense, it is the main
object in any gauge field action. Mathematically, the nonabelian field
strength is recognized as the two-form curvature in a fiber bundle.
Surprisingly, the possibilities to make self-dual such a two-form is very
restrictive. In fact, independently of the signature of spacetime,
self-duality seems to work only in dimensions $1,2,4,$ and $8$. These
dimensions can be recognized as the dimensions associated with the
exceptional algebras: real, complex, quaternion and octonion algebras. Of
course, in order to make sense of self-duality in any dimension one may
introduce $p$-form gauge fields but this case works only as an abelian
theory [26]. If one is thinking about gravity as a nonabelian gauge theory,
then one obtains a similar conclusion concerning the dimensions $1,2,4,$ and 
$8$. Therefore, from this point of view it seems natural to consider the
self-dual supergravity in eight dimensions. And the best way to do this is
by using a superfield prescription with an octonion algebra as a background
structure. Pursuing this idea we first applied a superfield formalism to
self-dual supergravity in four dimensions showing the consistency of the
procedure and then we applied similar technique for the case of self-dual
supergravity in eight dimensions.

Eight dimensional self-dual supergravity has been previously studied by
Nishino and Rajpoot [27]-[29] (see also Refs. [24] and [30]-[31]) whom based
their motivation on twelve dimensional supergravity [32]. However, in this
formalism the superfield prescription is not used and there is not any guide
for a connection with the Ashtekar formalism. Nevertheless, it seems
interesting for further research to find a connection between the works of
these authors and the formalism presented here. Eleven-dimensional
supergravity in light-cone superspace [33] has been studied recently. The
formalism in this work is based on the decomposition $SO(9)\supset
SO(2)\times SO(7)$. The $SO(7)$ subgroup is used on the superfield formalism
without using the octonion structure. Perhaps, our work may be useful in
this direction.

It has been noted [34] that the $\varepsilon -$symbol $\varepsilon ^{\mu \nu
\alpha \beta }$ is a chirotope [35] and that $\eta ^{\mu \nu \rho \sigma }$
is related to the Fano matroid [36]. This means that the Lagrangians (76)
and (79) contain information of the oriented matroid theory [35]. Therefore,
the present work suggested that, besides all the relations between matroids
and different aspects of $M-$theory, including gravity [37], Chern-Simons
theory [38] and String theory [39], Matroid theory should be connected with
the superfield formalism.

Starting with the supersymmetric Lagrangian in eight dimensions given in
(79) one may be interested in studying self-duality structure in
supersymmetric $10+2-$dimensional model via the two projections $%
10+2\rightarrow (1+3)+(1+7)$ and $10+2\rightarrow (2+2)+(0+8)$. If this
program goes on the right route then we shall become closer to find a
connection between the "self-dual" supergravity theory and the
supersymmetric Ashtekar formalism, in $10+2-$dimensional spacetime.

Finally, in order to achieve a manifest $SO(3,2)$ symmetry in the context of
the bosonic MacDowell-Mansouri formulation Stelle and West [40] introduced
the idea of compensator field. In Ref. [41] Vasiliev developed the
MacDowell-Mansouri-Stelle-West action for any dimension and in Ref. [42] the
compensator approach was applied to the study of $N=1$ supergravity in four
dimensions as a gauge theory of $OSp(1|4).$ Thus, it seems interesting to
see whether the results of the present work may be combined with the idea of
a compensator approach in higher dimensional supergravity theories.

\smallskip\ 

\textbf{Acknowledgment: }I would like to thank H. Garc\'{\i}a-Compe\'{a}n,
O. Obreg\'{o}n and C. Ram\'{\i}rez for their helpful comments.

\end{document}